\documentstyle[aps,multicol,epsfig]{revtex}
\begin{document}
\draft

\title{Localized and Delocalized Charge Transport in Single-Wall
Carbon-Nanotube Mats}
\author{O. Hilt and H. B. Brom}
\address{
Kamerlingh Onnes Laboratory, Leiden University, P.O. Box 9504,
2300 RA Leiden, The Netherlands}
\author{M. Ahlskog}
\address{
Low Temperature Laboratory, Helsinki University of Technology,
FIN-02015 HUT, Finland}

\date{November 16, 1999, accepted for Rapid Communications}
\maketitle

\begin{abstract}
We measured the complex dielectric constant in mats of single-wall
carbon-nanotubes between 2.7 K and 300 K up to 0.5 THz.  The data are well
understood in a Drude approach with a negligible temperature dependence of
the plasma frequency $\omega_p$ and scattering time $\tau$ with an additional
contribution of localized charges.  The dielectric properties resemble those
of the best ''metallic'' polypyrroles and polyanilines. The absence of
metallic islands makes the mats a relevant piece in the puzzle of the
interpretation of $\tau$ and $\omega_p$ in these polymers.
\end{abstract}


\begin{multicols}{2}
\settowidth{\columnwidth}{aaaaaaaaaaaaaaaaaaaaaaaaaaaaaaaaaaaaaaaaaaaaaaaaaa}

Depending on the wrapping of the graphene sheet, the intrinsic
electronic properties of single-wall carbon-nanotubes (SWNTs) are
either semiconducting (zigzag and most chiral nanotubes) or metallic
(armchair and part of the chiral nanotubes)\cite{Hamada92,Wildoer98}.  Single
ropes of armchaired SWNTs as well as entangled networks (mats) show a
decrease of the dc conductivity, $\sigma_{\rm dc}$, with increasing
temperature ($T$) for $T$ above a critical temperature
($T^*$)\cite{Fischer97,Kane98,Kaiser98,Petit97}. For $T<T^*$,
$\sigma_{\rm dc}$ decreases with cooling. $T^*$ typically lies
between 40~K and 250~K and depends on the morphology and the degree
of disorder\cite{Kane98}. The transition from $d\sigma_{\rm dc}/dT
<0$ to $d\sigma_{\rm dc}/dT > 0$ at $T^*$ is ascribed to structural
defects and built-in impurities of the individual
nanotubes\cite{Fuhrer99,Lee97}, or to barriers between the nanotubes or
ropes limiting the extension of the charge-carrier states\cite{Kaiser98}.
Additional information about this issue can be obtained from frequency-
dependent phase-sensitive permittivity experiments, giving the complex
conductivity $\sigma(\omega)= \sigma'(\omega) + i\sigma''(\omega)$ or
dielectric constant, $\epsilon(\omega)=\epsilon'(\omega) -
i\epsilon''(\omega)$\cite{sigma}.  Different frequency dependencies of
$\sigma'$ and $\epsilon'$ are expected for the two limiting cases of
localized and delocalized charge transport and also the sign of $\epsilon'$
changes with the two models. Here we apply this method to mats of SWNTs. The
data do not need Kramers-Kronig analysis, which is a great advantage for the
analysis of the response of the delocalized charge carriers\cite{Chapman99}.
Like heavily doped polymers SWNTs are shown to be an example of a system with
exceptional long scattering times $\tau$ and low plasma frequencies
$\omega_p$, but in contrast to the polymers metallic islands can be excluded
as an explanation.  For that reason SWNTs form an important new element in
the not-yet understood physics behind $\tau$ and $\omega_p$.

Weak inter-nanotube contacts or strong intra-nanotube defects might lead to
charge localization.  The motion of the localized charge carriers will be
diffusion controlled and the frequency-dependent conductivity
$\sigma(\omega) = ({ne^2}/{k_B T})D(\omega)$ is given by linear response
theory as\cite{Scherlax73}
\begin{equation}
D(\omega) = -\frac{1}{2d}\: {\omega}^2 \int_{0}^{\infty}
\langle (r(t) - r(0))^2 \rangle e^{-i\omega t} dt
\end{equation}
with $d$ the dimensionality of the transport system, $r$  the
charge-carrier position and $\langle \rangle$ the configurational average.
Eq. (1) reduces to Fick's law for a frequency independent $D$.  With
increasing frequency $\sigma'$ will increase while the positive $\epsilon'$
decreases\cite{Schirmacher91,Adriaanse97}.  In the delocalized case
charge transport is expected to follow the scheme of a Drude electron-gas
with a conductivity:
\begin{equation}
\sigma(\omega) = \epsilon_0\:\omega_p^{\:2} \tau
\:\frac{1}{1+i\omega\tau}
\end{equation}
with $\omega_p^2 = {nq^2}/({\epsilon_0 m^*})$\cite{AM}.
For most metals $\omega_p \sim 10^{15}$ s$^{-1}$.  If charge transport is
governed by (anomalous) diffusion (Eq.(1)), $d\sigma'/d\omega \ge 0$ and
$\epsilon' \ge 1$. In contrast, the Drude model of delocalized charge
transport (Eq.(2)) predicts $\epsilon' < 0$ for $\omega\ll\omega_p$ and
$d\sigma'/d\omega \le 0$ and $d\epsilon'/d\omega \ge 0$ for $\omega \sim
\tau^{-1}$.

SWNT mats\cite{Rice99} were prepared by vacuum-filtering a suspension of
SWNT's in water with approx. 0.5 \% Triton X-100, a non-ionic surfactant,
through filter paper with a pore size of 1 $\mu$m. The SWNT's were
produced using laser-ablation\cite{Thess96}. Purification of the
mats was performed by washing the filter paper with the attached
SWNT mat with deionized water to remove the Triton X-100 and with
methanol to remove residual NaOH\cite{Rinzler98}. In this way,
mats with a diameter of 34 mm and a thickness of typically 10
$\mu$m were obtained. Some of these SWNT mats were investigated
with the filter paper attached to it. Other mats could be peeled
off the filter paper. Up to now SWNT mats are always mixtures of chiral,
zig-zag and armchaired nanotubes. The fraction of metallic nanotubes is
estimated to be of the order of 0.1-0.5\cite{Fischer97,Petit97,Ugawa99}.  For
these strand-like materials such numbers are sufficiently high to be well
above the percolation threshold for dc-conduction\cite{Adriaanse97}.

The dc conductivity ($\sigma_{\rm dc}$) of the mats was measured with the
four-probe technique. The $T$-dependent dc measurements were performed
in an Oxford flow (down to 4 K) and a $^3$He cryostat (down to 0.4
K). Complex $\sigma(\omega)$ or $\epsilon(\omega)$-data in the GHz
regime were obtained by running-wave transmission-measurements with the
electric-field vector parallel to the plane of the mat. The transmission and
the phase shift introduced by the sample were directly measured with an ABmm
millimeter-wave network analyzer.  The complex dielectric constant at the
given $\omega$ could be obtained by fitting the data to first principles
formulae\cite{Reedijk99} without the need of a Kramers-Kronig analysis.  The
attenuation of the samples was approximately 45 dB. For the transmission
experiments at 10 GHz and at 15 GHz, resp.  X-band and P-band rectangular
waveguides were used.  The sample was mounted onto a choke flange of the
waveguides.  Between 40 GHz and 500 GHz a free-space electromagnetic wave,
focused with polyethylene lenses was transmitting the SWNT mat. For the
$T$-dependent measurements at 285 GHz the SWNT mat was mounted in an optical
He-flow cryostat. The mm-wave reached the sample by passing two quartz and
capton windows with a diameter of
40 mm. The sample space of the cryostat was filled with liquid He
for measurements below 4.2 K. The resulting change of $\epsilon$ of the
medium surrounding the sample was considered in the analysis.\\

\begin{figure}[htb]
\begin{center}
\leavevmode
\epsfig{figure=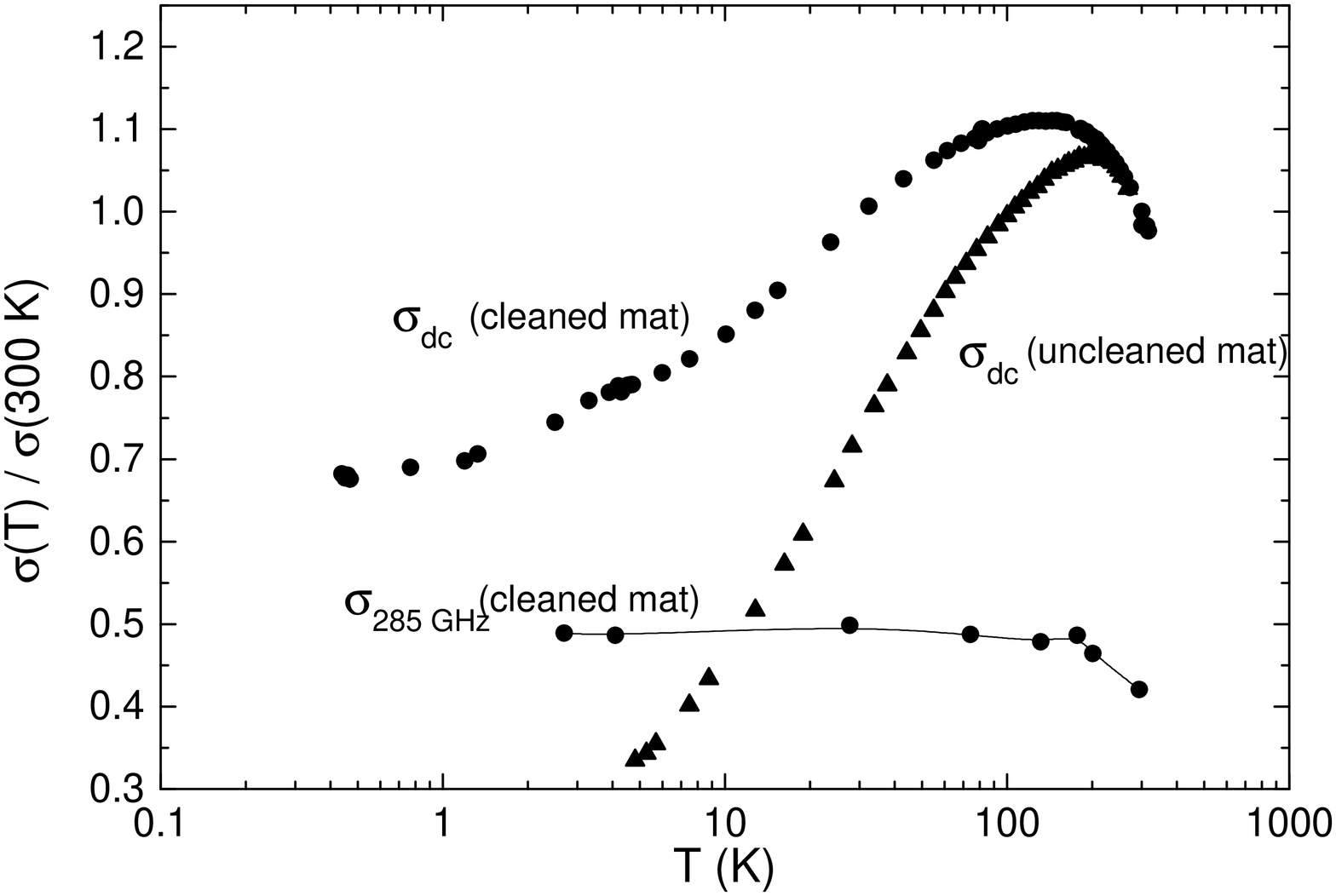,height=7cm,angle=-90}
\end{center}
\caption{Normalized $T$ dependence of $\sigma_{\rm dc}$
(circles) and $\sigma'_{280 {\rm GHz}}$ (open squares) for a cleaned SWNT
mat. The triangles show $\sigma_{\rm dc}(T)$ for a desiccated mat.}
\label{fig1}
\end{figure}

\begin{figure}[htb]
\begin{center}
\leavevmode
\epsfig{figure=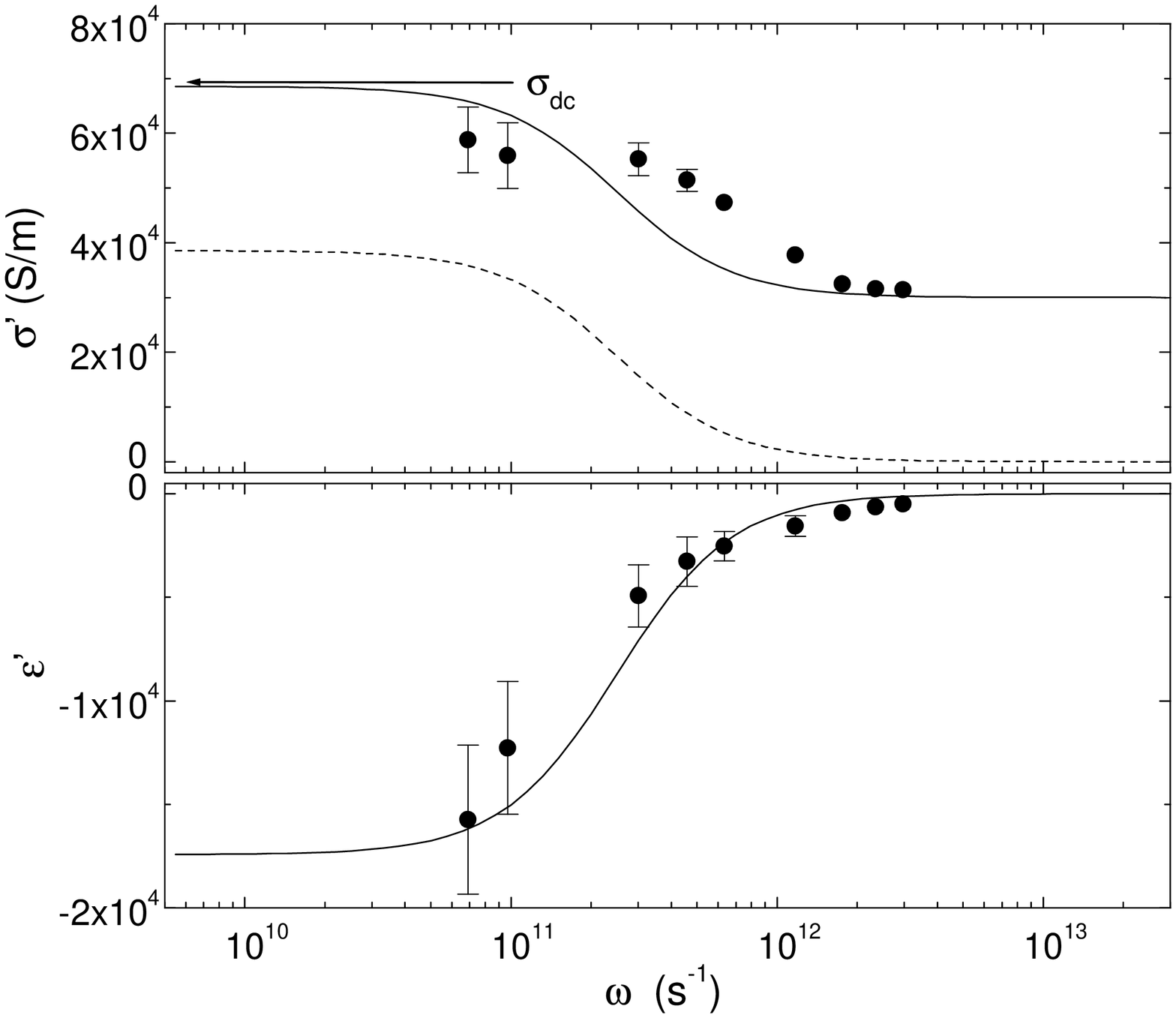,height=7cm,angle=-90}
\end{center}
\caption{Dependence on $\omega$ of $\epsilon'$ and $\sigma'$ for cleaned SWNT
mats at 300~K (full circles). The arrow marks $\sigma_{\rm dc}$.
Lines are fits with a Drude model ($\tau= 4.0$ ps and $\omega_p = 3.3 \times
10^{13}$ s$^{-1}$) without (dashed line) and with (full line) background
conductivity $\sigma_b = 3\times10^4$ S/m.  }
\label{fig2}
\end{figure}

\begin{figure}[htb]
\begin{center}
\leavevmode
\epsfig{figure=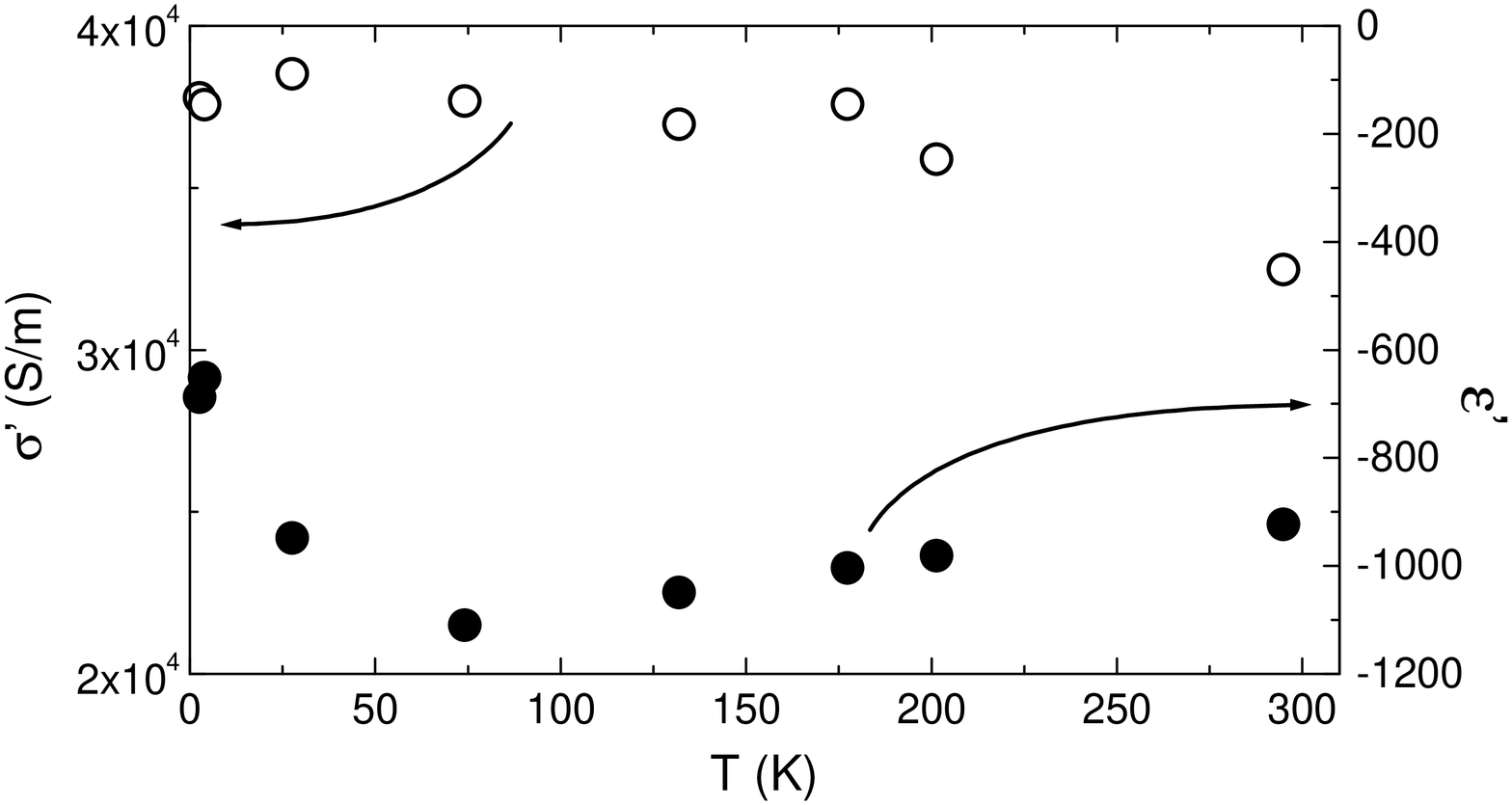,height=7cm,angle=-90}
\end{center}
\caption{$T$ dependence of $\sigma'$ (open circles)and $\epsilon'$ (full
circles) for cleaned SWNT mats between 2.7 K and 300 K at 285 GHz.}
\label{fig3}
\end{figure}

In Fig.~\ref{fig1} the $T$ dependences of $\sigma'_{\rm 285 GHz}$ and
$\sigma_{\rm dc}$, normalized to the 300~K value of $6.9\times 10^4$~S/m for
the cleaned and $2\times 10^4$~S/m for the uncleaned SWNT mats are shown.
Clearly visible is the change from a positive $d\sigma'/dT$ at low
$T$ to a negative $d\sigma'/dT$ at high $T$. The transition temperatures
between the two regimes is close to $T^*= 120$~K in agreement with recently
reported values for similar prepared SWNT mats\cite{Fischer97} (for the
uncleaned mat $T^* \sim 200$~K).  The measured frequency dependences of
$\epsilon'$ and $\sigma'$   show the following main features, see
Fig~\ref{fig2}: the conductivity almost keeps its dc value up to about 10 GHz
and decreases at higher frequencies. Saturation is observed close to 1
THz. The dielectric constant increases from $\epsilon' \sim -10^4$
in the 10 GHz-regime to $\epsilon' \sim -100$ close to 1 THz, apparently
approaching the regime of $\epsilon'> 0$ at still higher frequencies.
The dielectric constant at 285 GHz remains negative down to 2.7 K, see the
temperature-dependent data in Fig.~\ref{fig3}. For $T > T^*$, $\sigma'_{285
{\rm GHz}}(T)$ is proportional to $\sigma_{\rm dc}(T)$.  For $T < T^*$,
$\sigma'_{285 {\rm GHz}}(T)$ remains constant, while $\sigma_{\rm dc}(T)$
decreases with decreasing $T$, see Fig.~\ref{fig1}.

\begin{figure}[htb]
\begin{center}
\leavevmode
\epsfig{figure=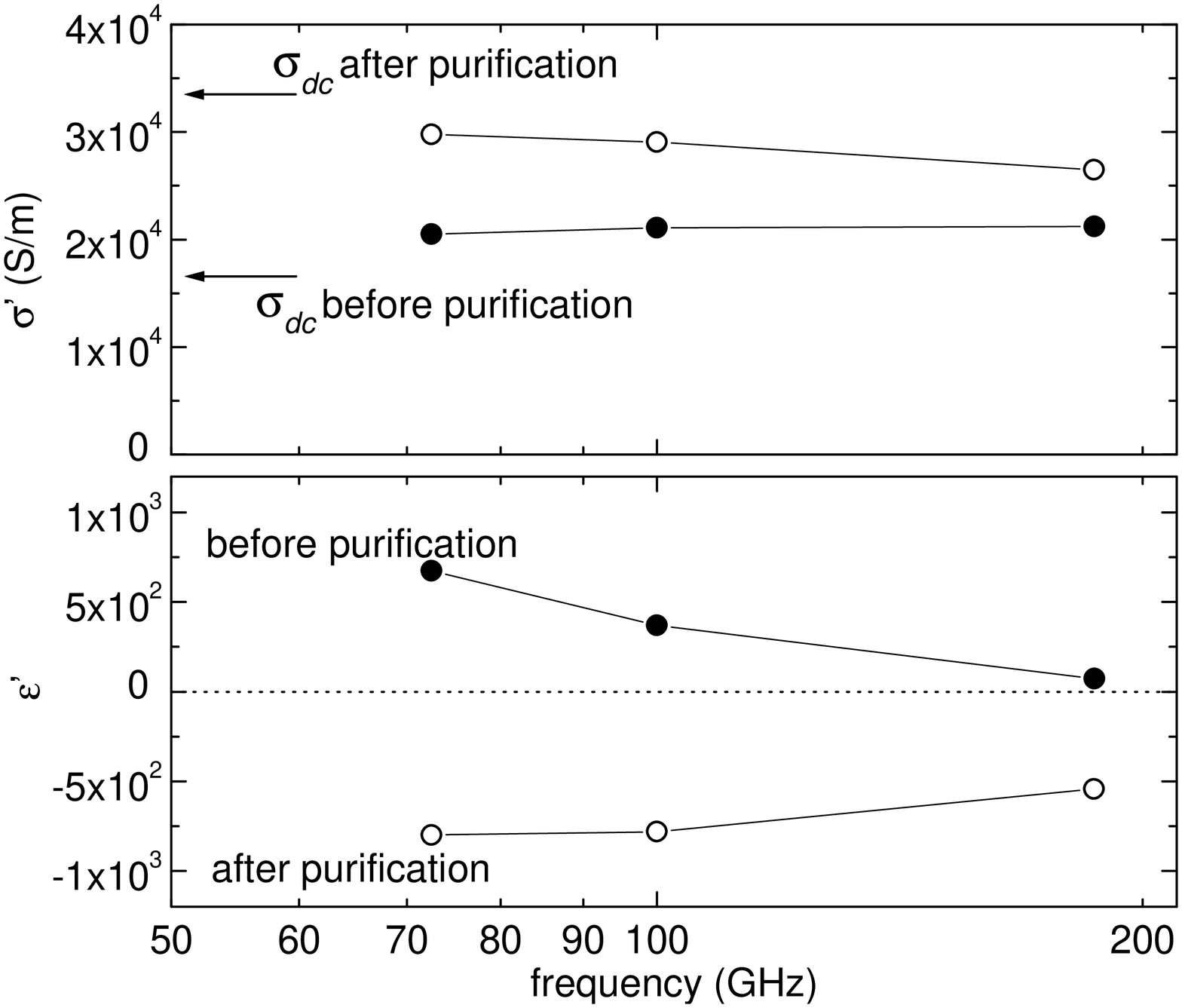,height=7cm,angle=-90}
\end{center}
\caption{Frequency dependence of $\sigma'$ and $\epsilon'$ for a desiccated
SWNT mat before (full circles) and after (open circles)
purification at 300~K. The charge transport is dominated by localized
carriers in the unpurified mat and shows metallic properties after
purification.}
\label{fig4}
\end{figure}

To study the influence of inter-rope barriers, desiccated SWNT mats
were prepared by evaporating the water after casting the
suspension onto a quartz substrate. After desiccation, the remaining
surfactant covers the SWNT-ropes. Fig.~\ref{fig4} shows the
dielectric constant to be positive for such an unpurified SWNT mat.
After rinsing the mat with deionized water and methanol, the
charge-transport properties turned from a localized-carrier
dominated regime with $\epsilon' > 0$, $d\epsilon' / d\omega < 0$
and $d\sigma' / d\omega > 0$ for the unpurified sample to a
delocalized-carrier dominated (metallic) regime with an increased
$\sigma_{\rm dc}$, $\epsilon' < 0$, $d\epsilon' / d\omega < 0$ and
$d\sigma'/ d\omega < 0$, see Fig.~\ref{fig4}.

The peculiar $T$ dependence of $\sigma_{\rm dc}$ has been explained
as a transition from metallic charge-transport at high temperatures
to a non-metallic regime at low temperatures caused by
charge-transport barriers of the order of some 10 meV\cite{Kane97}. Such a
model is expected to give a thermally activated behavior of the conductivity
at very low temperatures.  In our data (see Fig.~\ref{fig1})
$\sigma_{\rm dc}(T)$ flattens below 10 K showing the model to be too crude.
The metallic nature at room temperature is supported by the
negative dielectric constant obtained from the high-frequency
dielectric measurements, and $d\sigma'/d\omega \le 0$ and
$d\epsilon'/d\omega \ge 0$ for $\omega \sim \tau^{-1}$, see Fig.~\ref{fig2}.
A qualitative discrepancy between the observed frequency dependencies and the
Drude behavior lies in the saturation of
$\sigma'$ for $\omega >\tau^{- 1}$. The high-frequency
conductivity remains at approx. 40 \% of $\sigma_{\rm dc}$. This
points to a background-conductivity, $\sigma_b$, due to localized
charge carriers present in the system. A possible increase with frequency
of this background\cite{Reedijk98} (here neglected) will enhance the Drude
response at the lowest frequencies at most by a factor of 2.  The positive
contribution to $\epsilon'$ (also neglected) is usually well below $10^3$ in
the GHz-regime and decreases with $\omega$.

A fit of the Drude model including background conductivity is shown
in Fig.~\ref{fig2}. At low frequencies the value of $\sigma'$ is
not accurate and the experimental error of $\epsilon'$ in the 10
GHz-range is considerable. Although discrepancies in the shape of
the frequency dependencies remain, the principal characteristics as
described above can be reproduced. The observed more stretched
frequency dependence is expected for a distribution of $\tau$ and
$\omega_p$. Given the experimental inaccuracies, the order of magnitude of
the fit parameters is correct.  The scattering time is estimated as $\tau = 2
- 5\times10^{-12}$ s, the plasma frequency as $\omega_p = 2.5 -
5.5\times10^{13}$ s$^{-1}$ and the background conductivity as
$\sigma_b = 2-3\times10^4$ S/m.

Using the Fermi velocity of graphite, $v_F = 8 \times 10^5$ m/s, the
scattering time $\tau$ gives a mean free path of $\Lambda \sim 3$ $\mu$m. A
similar value has been estimated from ESR measurements at 100~K on SWNT
mats\cite{Petit97}. From DC measurements on an isolated SWNT at a few mK
$\Lambda = 3$ $\mu$m is suggested as a lower limit\cite{Tans97}.  Although
theoretical arguments\cite{Farajian99} predict
a very low scattering probability with acoustic phonons inside a tube, which
might allow similar values also at higher temperatures, the $\Lambda$ found
here refers to 3D transport, for which such a value seems (far) too high. We
will return to this problem below.

The obtained plasma frequency is about one percent of $\omega_p$ for normal
metals. Assuming $m^* = m_e$, the value of $\omega_p$
implies a charge-carrier density $n = 3 - 9 \times 10^{23}$
m$^{-3}$. Correcting for the lower density of the mats ($\sim 0.65$
g/cm$^3$) compared to that of a SWNT ($\sim 2$ g/cm$^3$) and
assuming a fraction of $50\%$ metallic tubes\cite{Petit97} would
give $n \sim 4 \times 10^{24}$ m$^{-3}$, comparable to graphite.
However, for SWNT $n$ is predicted to be about $10^2 \times$
higher\cite{Mintmire92}. Fischer {\sl et al.}\cite{Fischer97} found after
mass-density corrections a more than $10 \times$ higher
$\sigma_{\rm dc}$ for single ropes than for mats. It is
plausible that in a mat only a fraction of the charge carriers
present participate in the delocalized (metallic) charge transport, while due
to localization the remaining charge carriers have a
smaller contribution to $\sigma_{\rm dc}$\cite{Bezryadin98}. In the model
proposed here localized charge carriers are incorporated and contribute to
$\sigma_b$.

At $T=300$~K $\sigma'_{\rm 285 GHz}$ is mainly due to $\sigma_b$, i.e.
localized charge carriers (see fit in Fig.2). At such a high frequency
$\sigma'(\omega)$ might well be determined by photon- in stead of
phonon-assisted hopping\cite{Boettger85,Reedijk98}, which explains the
constant value of $\sigma'_{\rm 285 GHz}(T)$ for $T \le 10^2$ K .
However, $\epsilon'$ is still dominated by the delocalized charges over the
whole temperature range ($\epsilon'<0$ at 285 GHz, Fig.~\ref{fig3}). The
decrease by 40 \% of $|\epsilon'|$ between 70 K and 2.7 K can be accounted
for by the decrease of carriers from the semiconducting tubes and a growing
contribution of localized states. It shows that the metallic part of
$\epsilon'(\omega)$ for $\omega>1/\tau$ has no strong $T$ dependence, which
implies an almost $T$ independence of $\omega_p$.  Also $\sigma_{\rm dc}$
at 0.4~K is only a factor 0.7 lower than at 300~K, which indicates
that not only $\omega_p$ but also $\tau$ have not changed appreciably with
$T$. An opening of a gap due to twistons in the order of 20 meV as suggested
by Kane and Mele\cite{Kane97} seems therefore unlikely.  Based on the
measured metallic low-temperature behavior down to 4.2 K of the
thermoelectric power in SWNT ropes, Hone {\it et al.}\cite{Hone98} also
excluded the opening of a gap at low $T$\cite{n1}.

The room temperature data presented in Fig.~\ref{fig4} confirm the importance
of inter-rope contacts at low temperatures. The purification procedure
removed the surfactant and other impurities from the surface of the
SWNT ropes allowing better contacts between ropes. Intra-tube or
intra-rope transport are likely not changed by the purification,
meaning that charge-localization effects due to defects in the
graphene-sheet pattern of the tubes or bending of them should be
unaffected. The effect of the purification on $\sigma_{\rm dc}(T)$, see
Fig.~\ref{fig1}, supports the picture.  The transition temperature $T^*$ for
the uncleaned sample is higher than for the cleaned one. Also, $d\sigma_{\rm
dc}/ dT$ below $T^*$ is larger for the latter. Both indicate that inter-rope
barriers limit $\sigma_{\rm dc}$ at low temperatures. These findings are
consistent with the higher $\sigma_{\rm dc}$, lower $T^*$ and
weaker $d\sigma_{\rm dc}/ dT$ below $T^*$\cite{Fischer97} of
single rope data, where inter-rope barriers are eliminated.

Highly conducting polymers like doped polyaniline (PAN) and
polypyrrole (PPy)\cite{Lee95,Kohlman97} and SWNT-mats show analogous
dielectric behavior. In the metallic polymers $\sigma_{\rm dc}$ typically has
a maximum value of order $10^4$ S/m (around 200~K) and decreases to lower
temperatures.  The values of $\epsilon'(\omega)$ are strongly frequency and
$T$ dependent.  Let us take one of the best conducting materials, PAN doped
with $d$,1-camphorsulfonic acid (PAN-CSA), as an example\cite{Kohlman97}.
Around 1 meV ($\omega \sim $ 1.5 THz) at 200 K $\epsilon'(\omega)$ is a few
times $-10^3$, and becomes less negative at lower $T$. For the same samples
at room temperature $\epsilon'(\omega)$ starts negative, becomes positive
around 30 meV, returns negative around 0.1 eV and finally comes close to zero
in the optical regime.  In the SWNT-mats the maximum value of $\sigma_{\rm
dc}$ is almost $10^5$ S/m, and $\epsilon$(285 GHz) reaches a value of $-10^3$
and decreases in absolute value with decreasing temperature (by a factor of
two at 4.2 K). At room temperature below 0.5 meV $\epsilon'(\omega)$ is
negative, likely becomes positive at higher energies and returns negative
again around 10 meV\cite{Ugawa99,note1,Chapman99}.  The comparison shows that
best conducting polymers are essentially behaving like well-rinsed mats of
single wall nanotubes (the same similarity exists between not-rinsed mats and
the slightly less conducting polymers).
Like in the mats, in these polymers values for $\Lambda$ will be of the order
of 100 nm for $v_F = 5 \times 10^5$ m/s\cite{Kohlman97}. In the metallic
polymers homogeneous and inhomogeneous disorder
models\cite{Kohlman97,Chapman99} are frequently used to explain these extreme
values. For the nanotubes crystalline regions are excluded (TEM pictures show
the nanotube mats to be completely entangled and disordered), which  at least
for the SWNT-mats requires an alternative for the inhomogeneous disorder
model.

In summary, we have shown that the delocalized properties of nanotube mats
can be completely determined by sub-THz measurements and are well described
in a Drude picture with a negligible temperature dependence of the plasma
frequency and scattering time.  The values of $\omega_p$ and $\tau$ resemble
those found in well conducting doped polymers.  When modeled with the usually
chosen Fermi velocities, unusually large values of the mean free path result
for both systems. This finding and the absence of crystalline regions in the
mats underline the need for a better description of these disordered
strand-like systems.

Acknowledgements. We like to acknowledge Hubert Martens for enlightening
discussions and critical reading of the manuscript and Roel Smit for support
in the experiments with the $^3$He cryostat.  This investigation is part of
the research program of FOM-PPM with financial support from NWO. M. Ahlskog
was  supported by the Academy of Finland.

\end{multicols}


\begin{thebibliography}{10}
\bibitem{Rice99}
{The nanotube-solutions were provided by Tubes@Rice. }
\end{thebibliography}
\end{document}